\documentclass[a4j]{article}
\usepackage[top=25truemm,bottom=25truemm,left=25truemm,right=25truemm]{geometry}
\usepackage[english]{babel}
\usepackage{authblk}
\usepackage{subfigure}
\usepackage{wrapfig}
\usepackage{graphicx}

\title{Characteristic study of silicon sensor for ILD ECAL}
\author{Shusuke Takada, Hiroto Hirai, Kiyotomo Kawagoe, Yohei Miyazaki, Yuji Sudo, \\ Taikan Suehara, Hiroki Sumida, Tatsuhiko Tomita, Hiraku Ueno, Tamaki Yoshioka}

\affil{Kyushu University}
\date{}

\begin{document}
\maketitle

\begin{abstract}
Excellent jet energy measurement is important at the International Linear Collider (ILC) because most of interesting physics processes decay into multi-jet final states.
We employ a particle flow method to reconstruct particles, hence International Large Detector (ILD) needs high spatial resolution which can separate each particle in jets.
We study pixelized silicon sensors as active material of ILD Silicon electromagnetic calorimeter (SiECAL).
This paper reports studies of temperature and humidity dependence on dark current and response of laser injection.
\end{abstract}

\footnote[0]{Talk presented at the International Workshop on Future Linear Colliders (LCWS14), Belgrade, Serbia, 6-10 October 2014.}

\section{Introduction}
The International Large Detector(ILD) is a proposed detector for the International Linear Collider (ILC)~[1,2].
Figure 1 shows the ILD and its electromagnetic calorimeter (ECAL).
The ILD ECAL is a sampling calorimeter consisting of tungsten absorber and subdivided sensors.
As candidates for the sensor technologies three types of sensors are designed, which are silicon detector, scintillation detector and hybrid of the two.
Particle Flow Algorithm (PFA)~[3] is planned to used as analytical method in ILC.
In PFA, particles are detected at optimal detectors.
Charged particles, photons and neutral hadrons are detected in a tracking detector, an electromagnetic calorimeter and a hadron calorimeter, respectively.
To improve the performance of PFA, the sensors are finely segmented to achieve excellent particle separation in jets.
The pixel size of sensors has to be less than 1 cm${^2}$ to satisfy a requirement for jet energy resolution.
The details of current ECAL design of ILD are described in the ILD-DBD.

\begin{figure}[htbp]
  \begin{center}
    \includegraphics[width=11.5cm,bb=0 0 555 235]{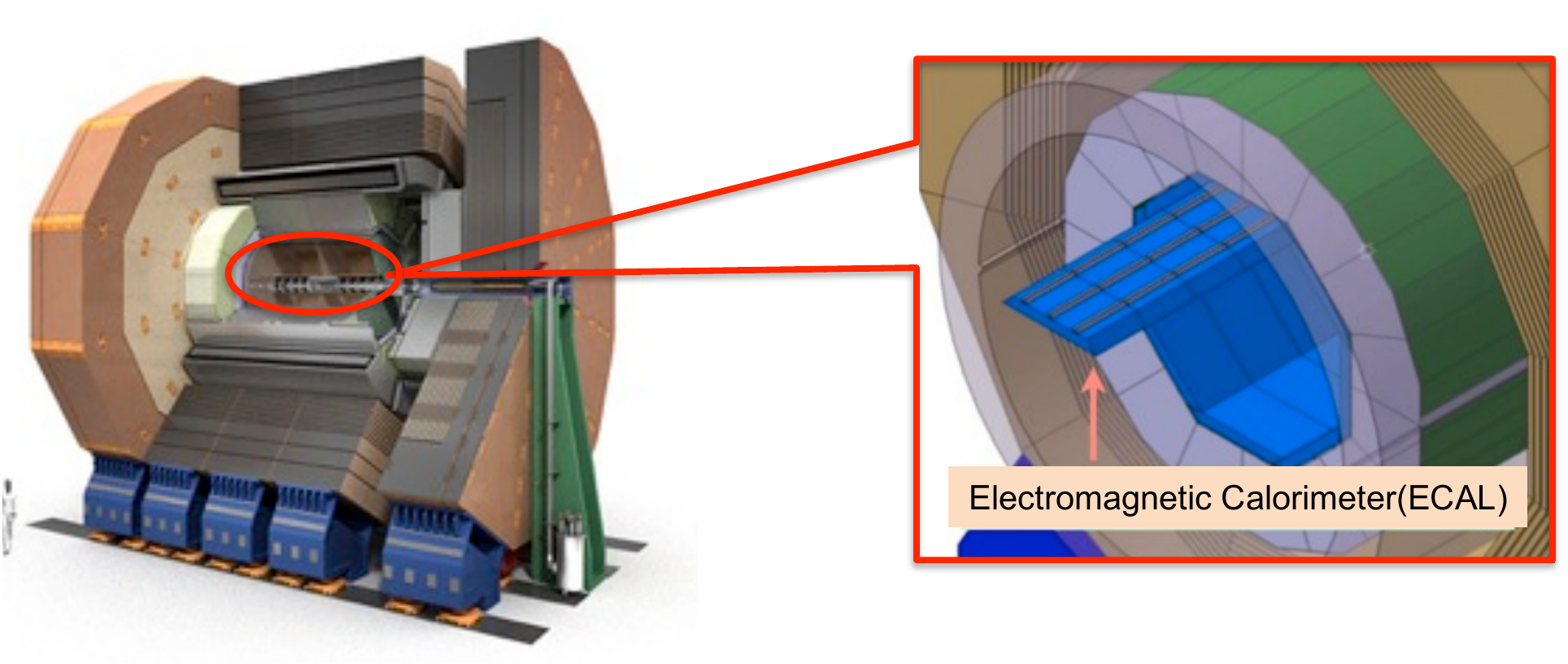}
    \caption{ILD detector and ECAL}
    \label{fig1}
  \end{center}
\end{figure}

\section{Pixelized silicon detector}
Figure 2 shows a sample of pixelized silicon sensor (Si-pad) which was made by Hamamatsu Photonics.
The specifications of Si-pad are the following: 5.5 $\times$ 5.5 mm${^2}$ pixel size, 320 $\mu$m thickness and 16 $\times$ 16 pixels.
A silicon sensor usually has ``Guard-ring" which is located at edge of the sensor.
Advantages of guard-ring are to collect surface current and to prevent an electric discharge at the sensor edge. The guard-ring, however, decreases sensitive area and makes crosstalk signal along the edge of the Si-pad.
Our motivation is to optimize Si-pad design by studying crosstalk caused by the guard-ring, crosstalk between pixels.
Dark current is measured as a basic property of the Si-pad and compared among the guard-ring types.

Figure 3 shows small Si-pad samples (baby chips) which are measured to compare the effects of different guard-ring designs.
Pixel size and thickness of the baby chips are the same as those of 16 $\times$ 16 pixels type.
Two types of baby chips were used.
One is  for comparison of guard-ring effects and the other is for measuring inter-pixel crosstalk.
For the guard-ring comparison we used 4 types of guard-ring(s), which are  0, 1, 2 and 4 guard-ring(s). Figure 3 and 4 show the structure of each guard-ring.
The guard-ring of the 1 guard-ring type is a continuous line without any divisions and those of the 2 and 4 guard-rings types are alternately divided.
Figure 5 shows a sample for inter-pixel crosstalk study.
Electrodes covering the chip is partly meshed in order to pass the laser photons though the mesh to the bulk of the silicon sensor. 

\begin{figure}[htbp]
  \begin{minipage}{0.5\hsize}
    \begin{center}
      \includegraphics[width=5cm, bb=0 0 842 596]{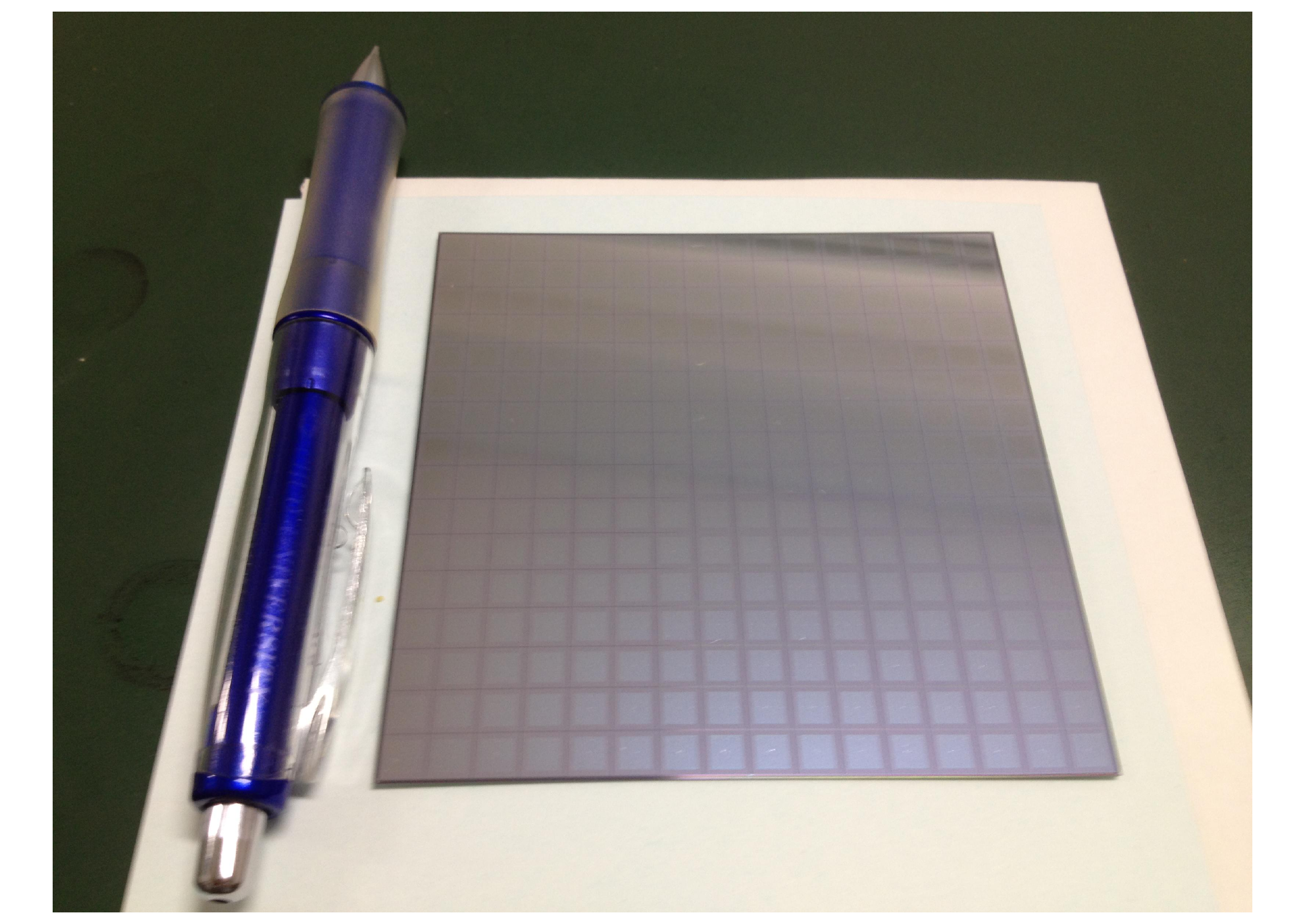}
      \caption{Picture of the 16 $\times$ 16 type Si-pad.}
      \label{fig2}
    \end{center}
  \end{minipage}  
  \begin{minipage}{0.5\hsize}
    \begin{center}
      \includegraphics[width=7.8cm,bb=0 0 842 596]{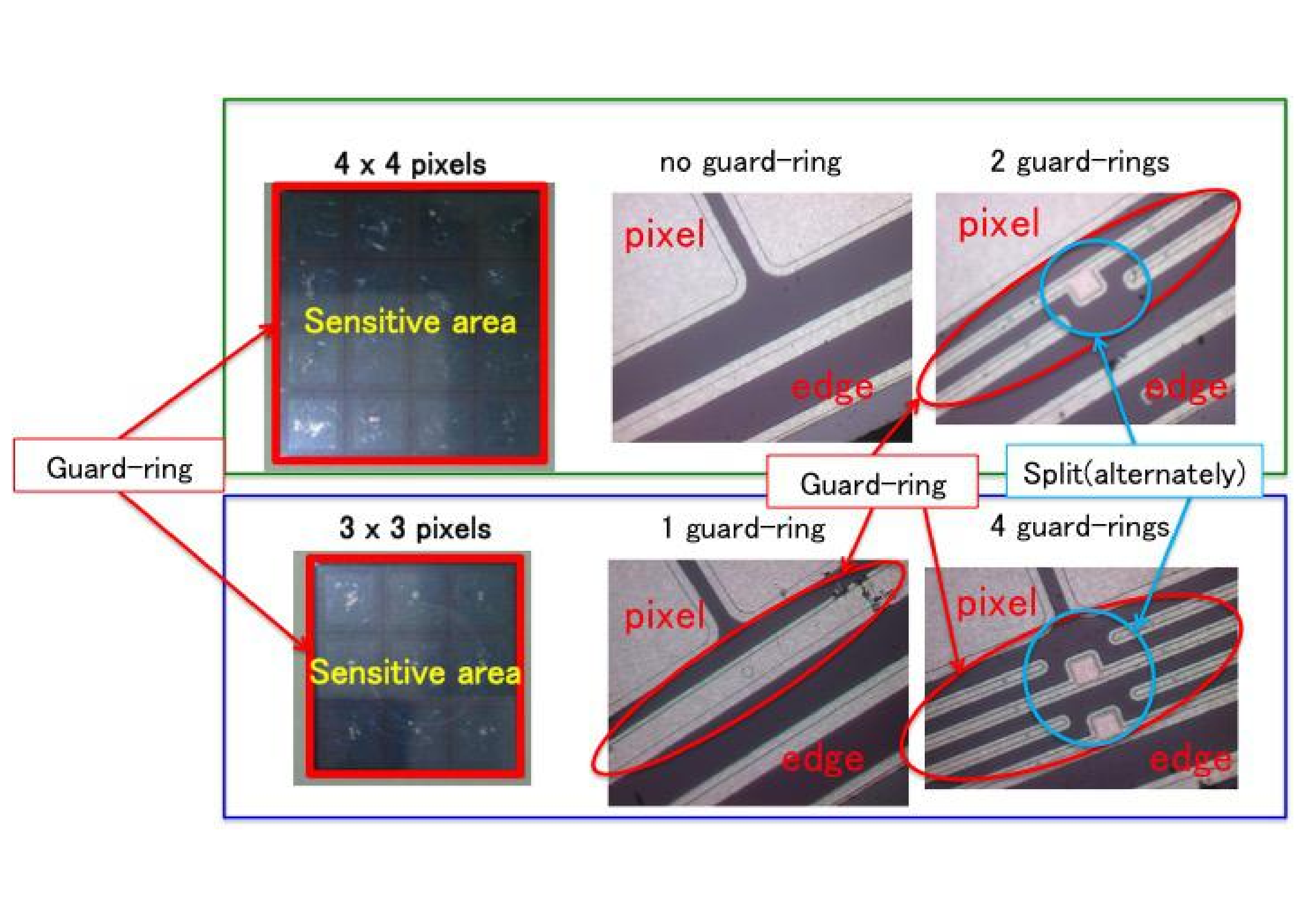}
      \caption{Picture of the baby chips. The right pictures show magnification images around the edge regions of the chips.}
      \label{fig3}
    \end{center}
  \end{minipage}
\end{figure}
\begin{figure}[htbp]
  \begin{center}
    \includegraphics[width=7cm, bb=0 0 665 418]{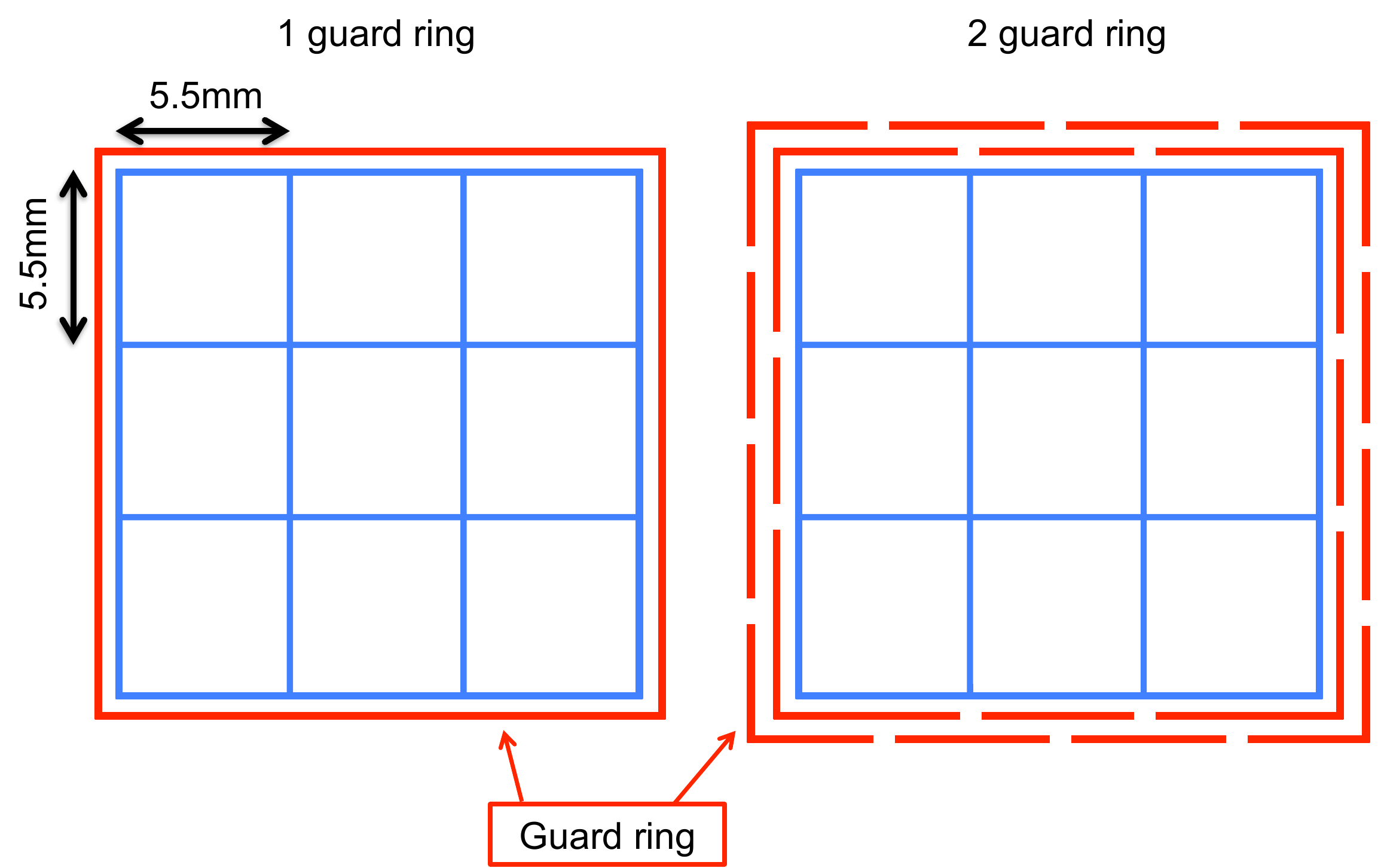}
    \caption{Schematics of the guard-rings of baby chips. Structure of the 4 guard-rings type is similar to the 2 guard-ring type.}
    \label{fig4}
  \end{center}
\end{figure}
\begin{figure}[htbp]
  \begin{center}
    \includegraphics[width=8.4cm, bb=0 0 842 596]{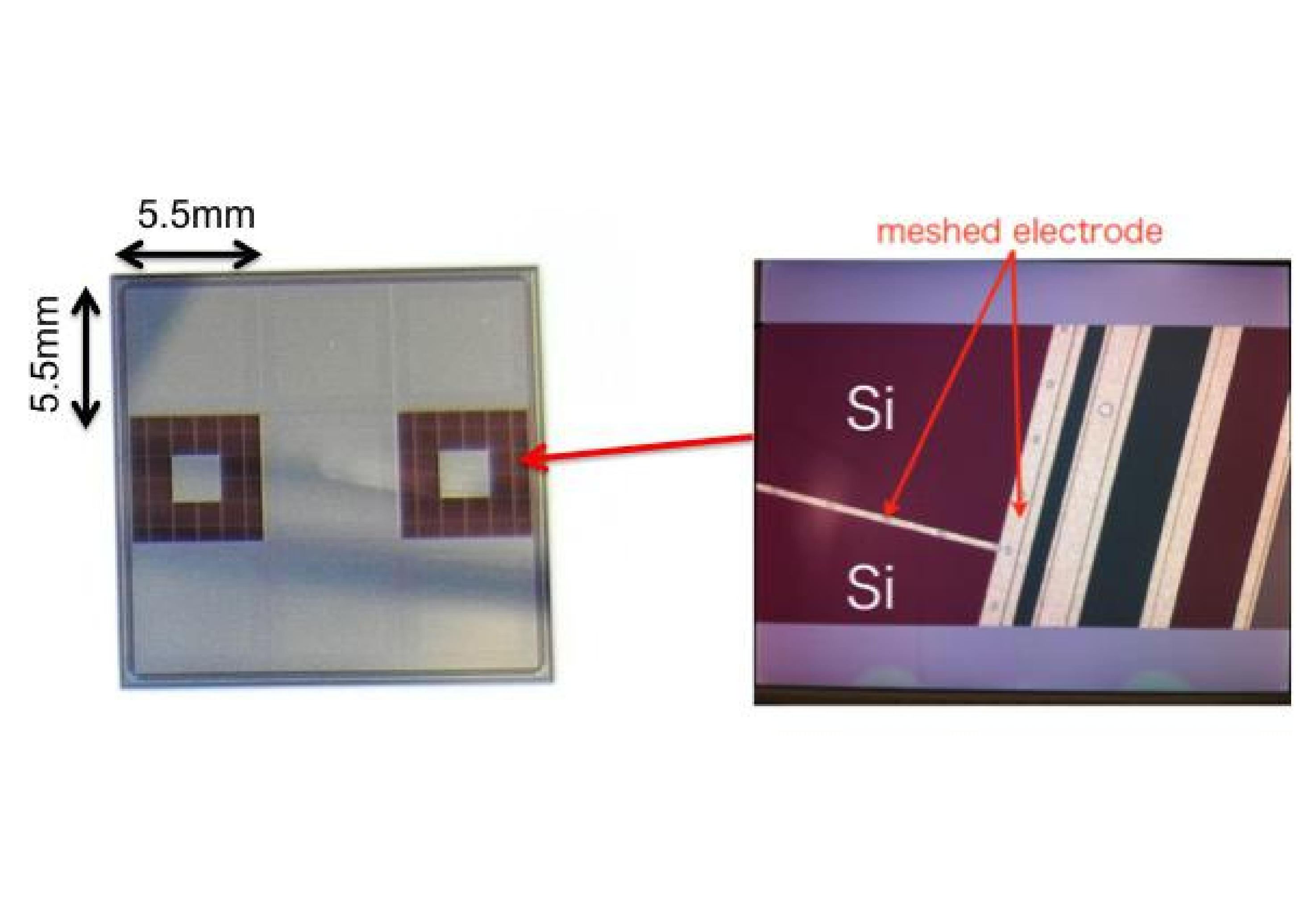}
    \caption{Pictures of the baby chip of meshed electrodes. The right picture is a magnification of the edge of one open pixel.}
    \label{fig5}
  \end{center}
\end{figure}

\section{Measurement systems}
We performed two measurements to optimize the Si-pad design.
One is the measurement of temperature and humidity dependence on dark current of Si-pad.
The other is the measurement of the response to laser injection to the edge of the baby chips and inside pixels of the chip.

\subsection{Setup of I-V measurement}
Figure 6 and 7 show the diagram and picture of the I-V measurement system.
Si-pad is put in a plastic box and readout pins touch each pixel.
Signal of readout pins are collected to a copper sheet and we measured summed current of Si-pad.
We put the measurement box in thermohygrostat to stabilize temperature and humidity.
During the measurement, we recorded temperature and humidity around Si-pad with a thermocouple and a humidity sensor.
\begin{figure}[htbp]
  \begin{minipage}{0.5\hsize}
    \begin{center}
      \includegraphics[width=6cm, bb=0 0 440 234]{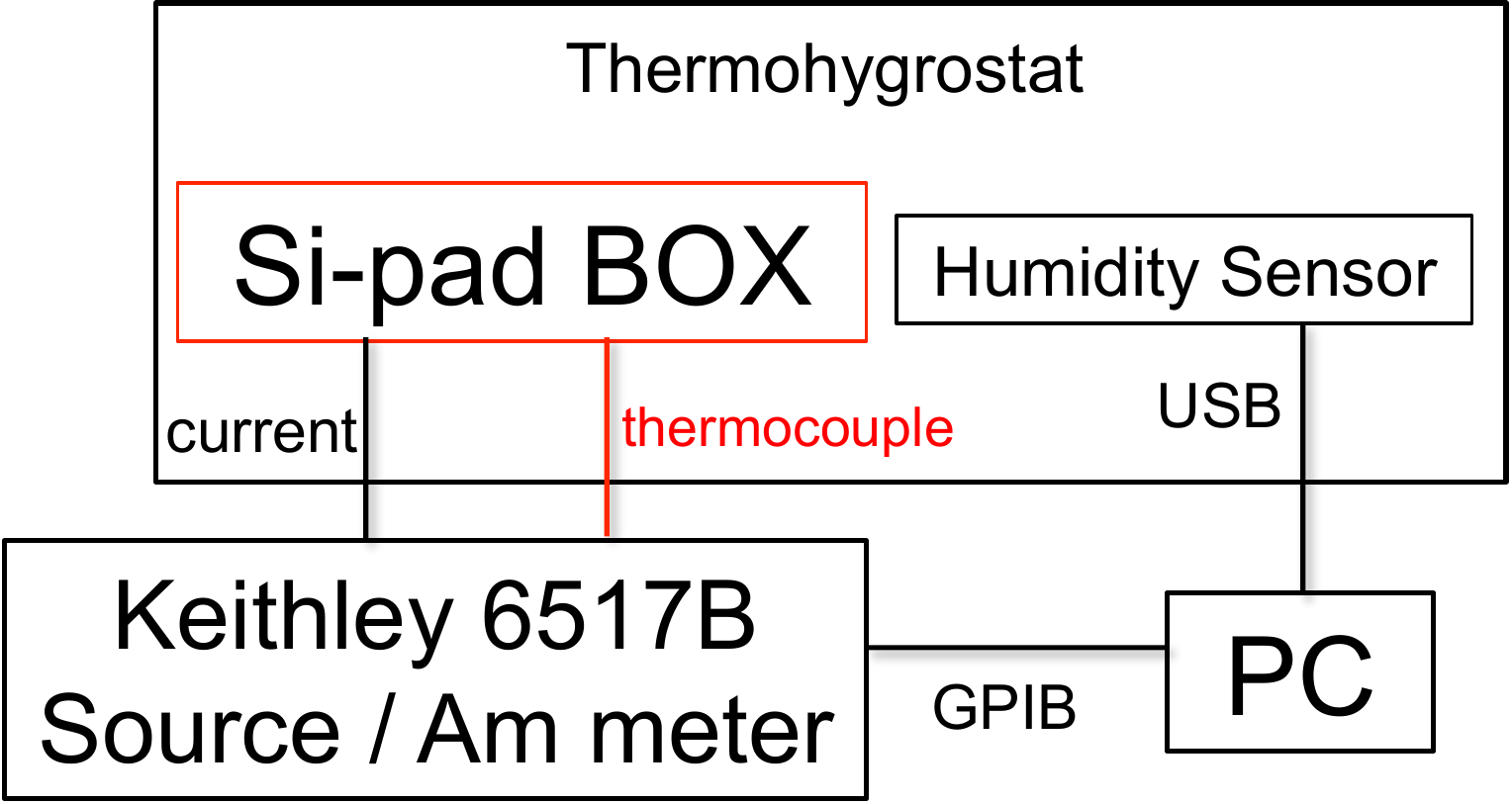}
      \caption{Schematic setup of I-V measurement. LabVIEW is used by data taking.}
      \label{fig6}
    \end{center}
  \end{minipage}
  \begin{minipage}{0.5\hsize}
    \begin{center}
      \includegraphics[width=6cm, bb=0 0 842 596]{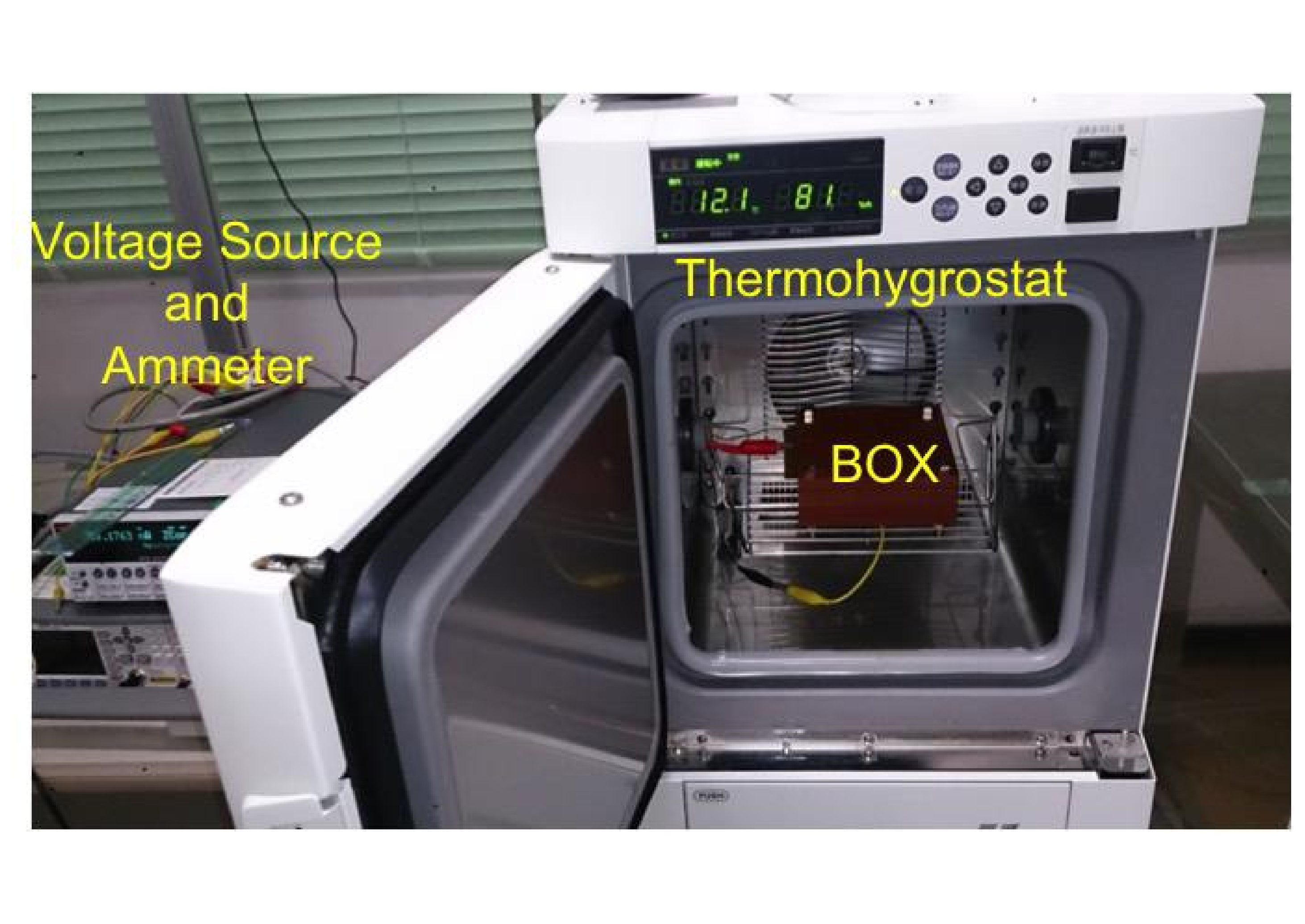}
      \caption{A picture of the setup of I-V measurement.}
      \label{fig7}
    \end{center}
  \end{minipage}
\end{figure}

\subsection{Setup of laser injection}
Figure 8 shows the setup of the measurement of laser injection.
Wavelength of the infrared laser is 1064 nm, which corresponds to 1.16 eV, slightly above the Silicon's band-gap energy of 1.12 eV so that one laser photon can produce an electron-hole pair.
The peak power of the laser is 13 kW and we can adjust the laser power by filters.
Reputation rate is 1 kHz. Laser spot size is less than 20 $\mu$m.
\begin{figure}[htbp]
  \begin{minipage}{0.5\hsize}
    \begin{center}
      \includegraphics[width=4cm, bb=0 0 596 842]{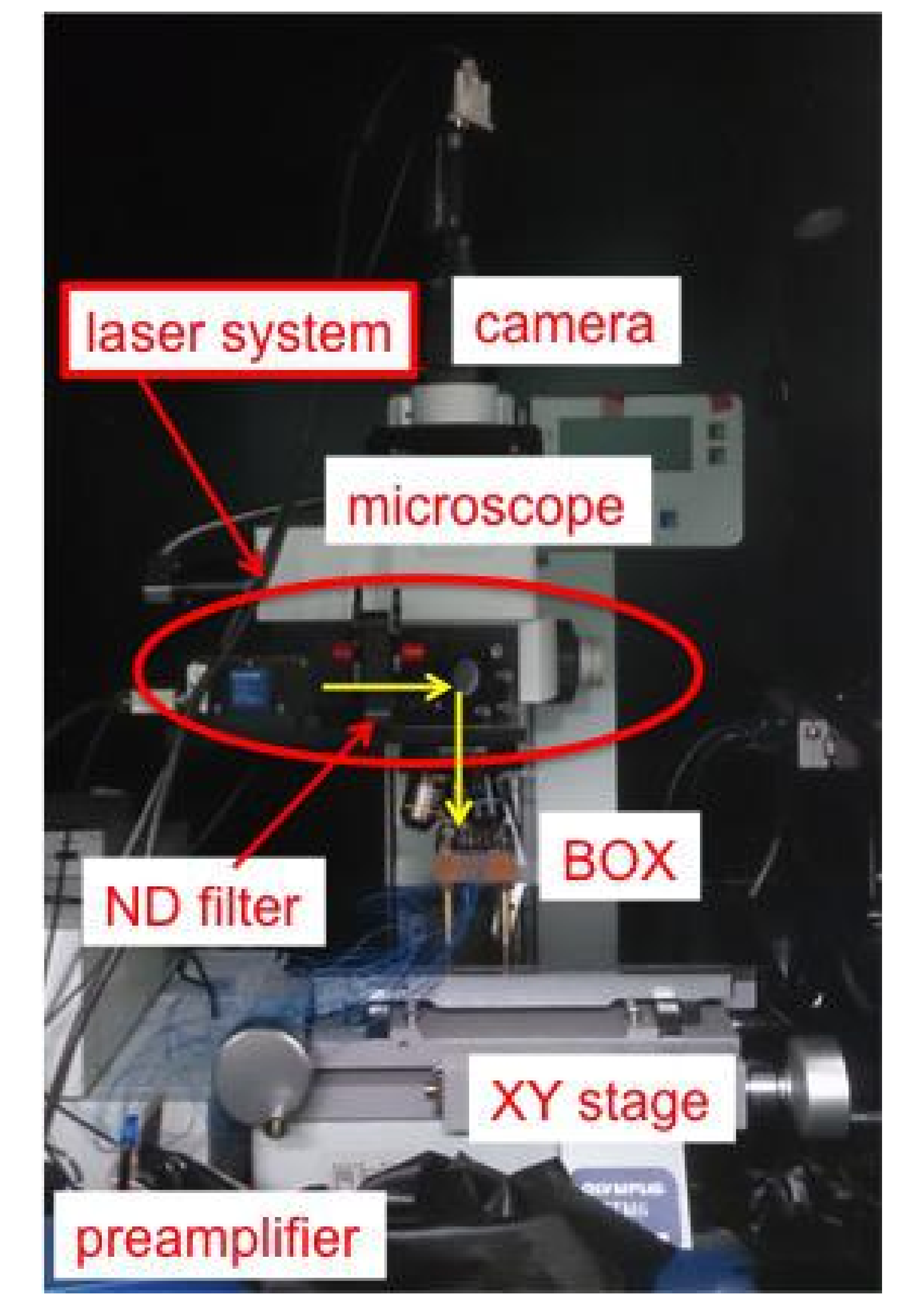}
      \caption{Picture of setup of Laser injection.}
      \label{fig8}
    \end{center}
  \end{minipage}
  \begin{minipage}{0.5\hsize}
    \begin{center}
      \includegraphics[width=6cm, bb=0 0 842 596]{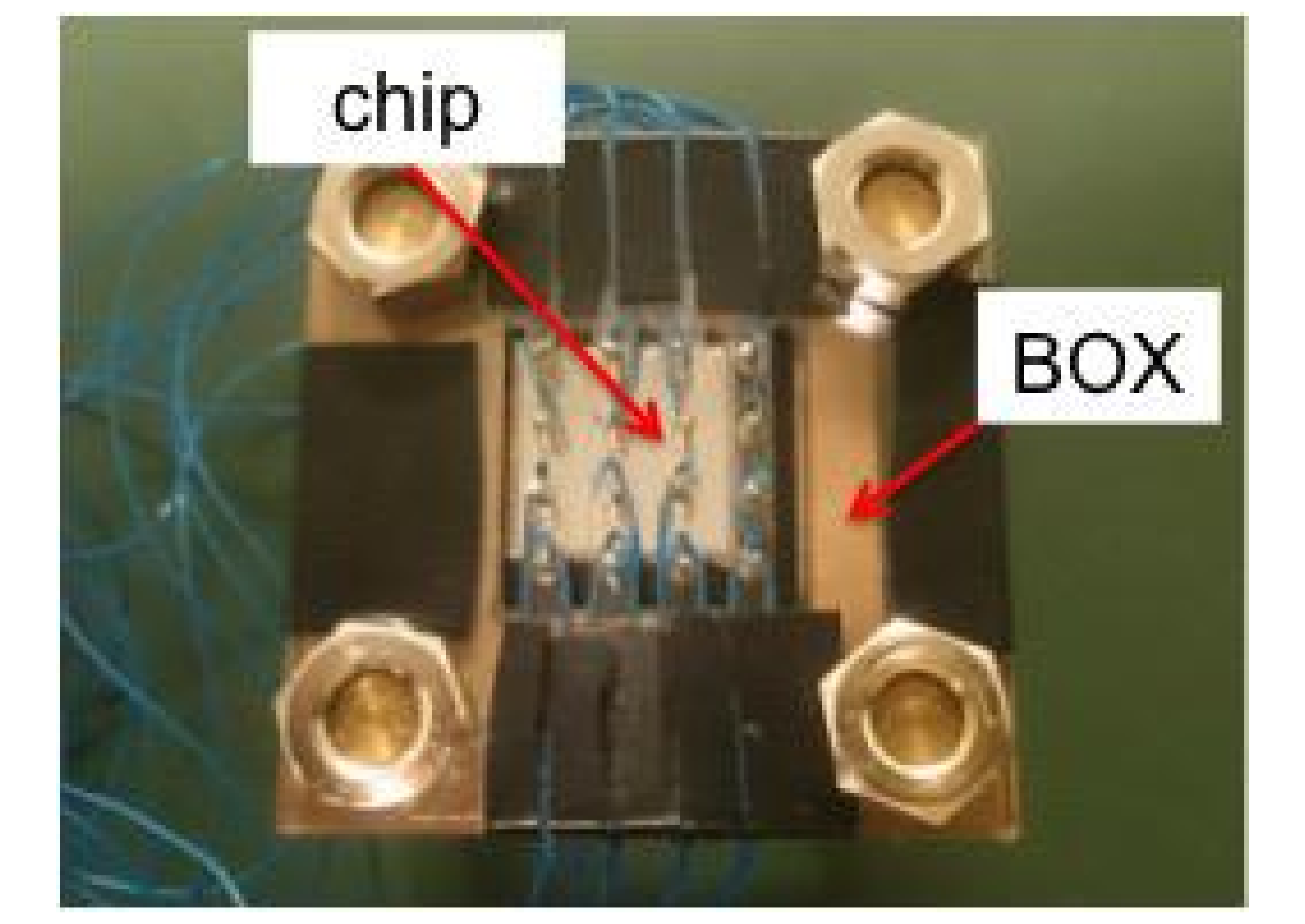}
      \caption{Picture of measurement box for laser injection. Length of the pins can be adjusted since the pin has a spring in it.}
      \label{fig9}
    \end{center}
  \end{minipage}
\end{figure}
Figure 9 shows a box for the laser injection. 
The readout pins touch each pixel and we can measure the signal from individual pixels.
We fix the readout pins by a thin acrylic plate, which has holes to inject laser light into pixels directly.
Injection point of the laser light is the gap region between the pixel edge and the guard-ring, because pixels are covered with aluminum electrode.
We prepared small Si-pads with meshed electrodes shown in Figure 10 to study crosstalk effect between pixels.
\begin{figure}[htbp]
  \begin{center}
    \includegraphics[width=12cm, bb=0 0 842 596]{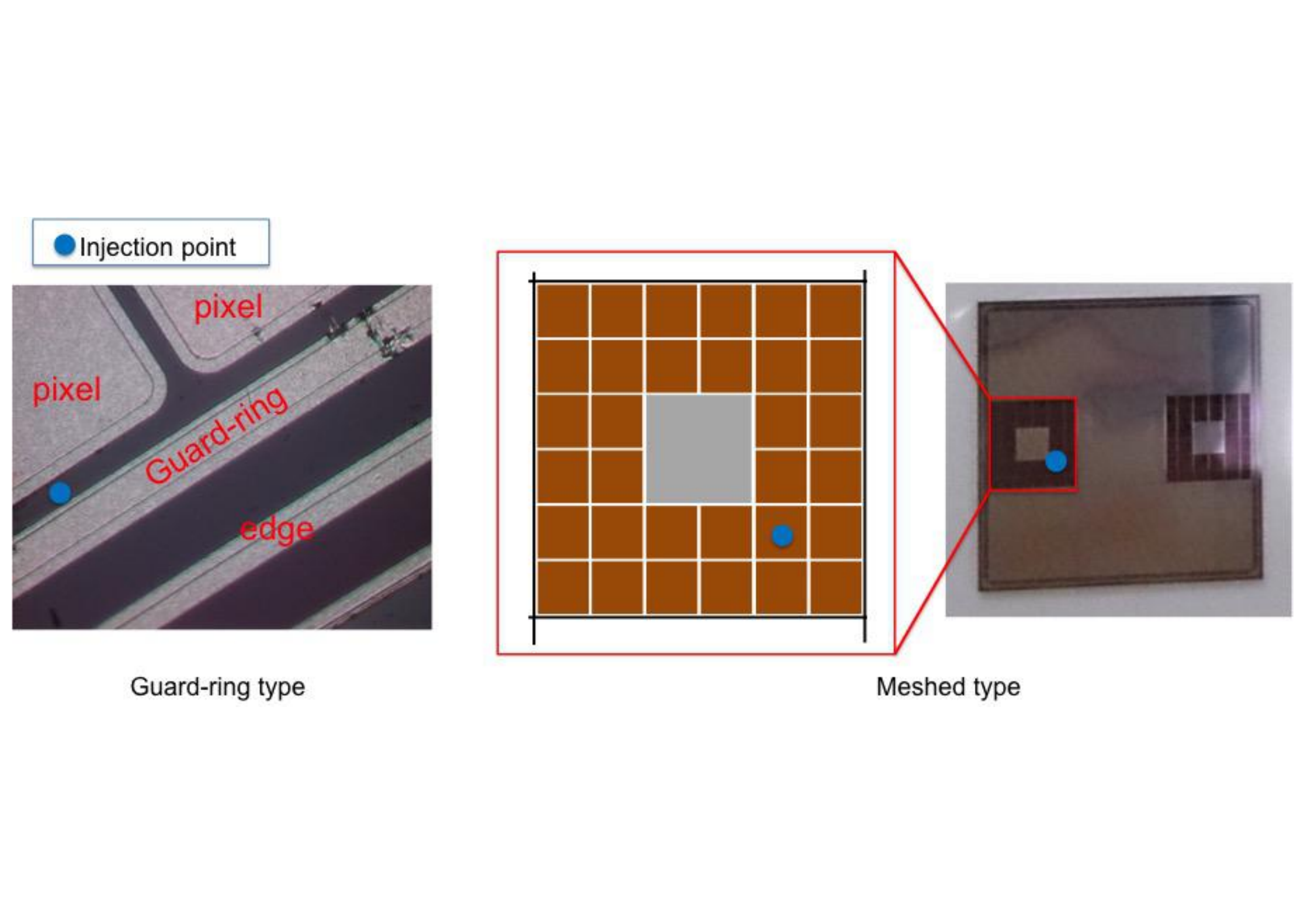}
    \caption{Points of laser injection for the comparison of guard-ring effect~(left) and for the inter-pixel crosstalk study with the meshed chip~(right). Blue dot shows the injection point.}
    \label{fig10}
  \end{center}
\end{figure}

\section{Results}
Figure 11 shows the result of temperature dependence of dark current.
There are no big differences among the guard-ring types.
Fitting function is expressed as
\begin{equation}
I(T)=AT^2{\rm exp}(-\frac{E_g(T)}{2k_BT})
\end{equation}
where $A$ is a constant factor, $T$ is temperature and $k_B$ is Boltzmann constant.
Silicon's band-gap energy $E_g$ also depends on temperature which is described by Varshni's empirical expression,
\begin{equation}
E_g(T)=E_g(0)-\frac{\alpha T^2}{\beta +T}
\end{equation}
where parameters $E_g(0)$, $\alpha$ and $\beta$ are approximately 1.1557 ${\rm [eV]}$, $7.021 \times 10^{-4}{\rm [eV/K]}$ and 1108 ${\rm [K]}$, respectively.
The function $E_g(T)$ has been experimentally determined~[4].
Current of edge region is larger than that of bulk region because electric field of edge of Si-pad differs from bulk region of it.
The fitting function, however, does not include the edge current.
Fitting result of silicon's band-gap energy is around 1.7 eV for all types of the guard-ring.
The big deviation from theoretical value of 1.1557 eV can be caused by edge current which is not considered in the current fitting.
This issue is under investigation.


However, we have a problem for reproducibility.
Figure 12 shows the result of temperature dependence of the 1 guard-ring type in multiple sets of measurements.
If once we take out the baby chip from measurement box and set it again, the dark current is significantly changed and the maximum difference is factor of two between different measurement sets.
The shift can be occurred by surface resistance of copper sheet whose condition is not good.
Figure 13 shows the result that the shift is reduced by clipping same area on the copper sheet.
It also have the shift, but it is much smaller than before.
Dark current of the baby chips, which is vertical axis, are normalized by size of the chips and constant term is added in fitting function.
At room temperature region, no significant difference was seen among different guard-ring types.
The band-gap energy is improved to 1.65 eV level.
At high temperature region, we can see a difference of chip size.
This should be also caused by the edge effect.

\begin{figure}[htbp]
  \begin{minipage}{0.5\hsize}
    \begin{center}
      \includegraphics[width=8.3cm, bb=0 0 605 362]{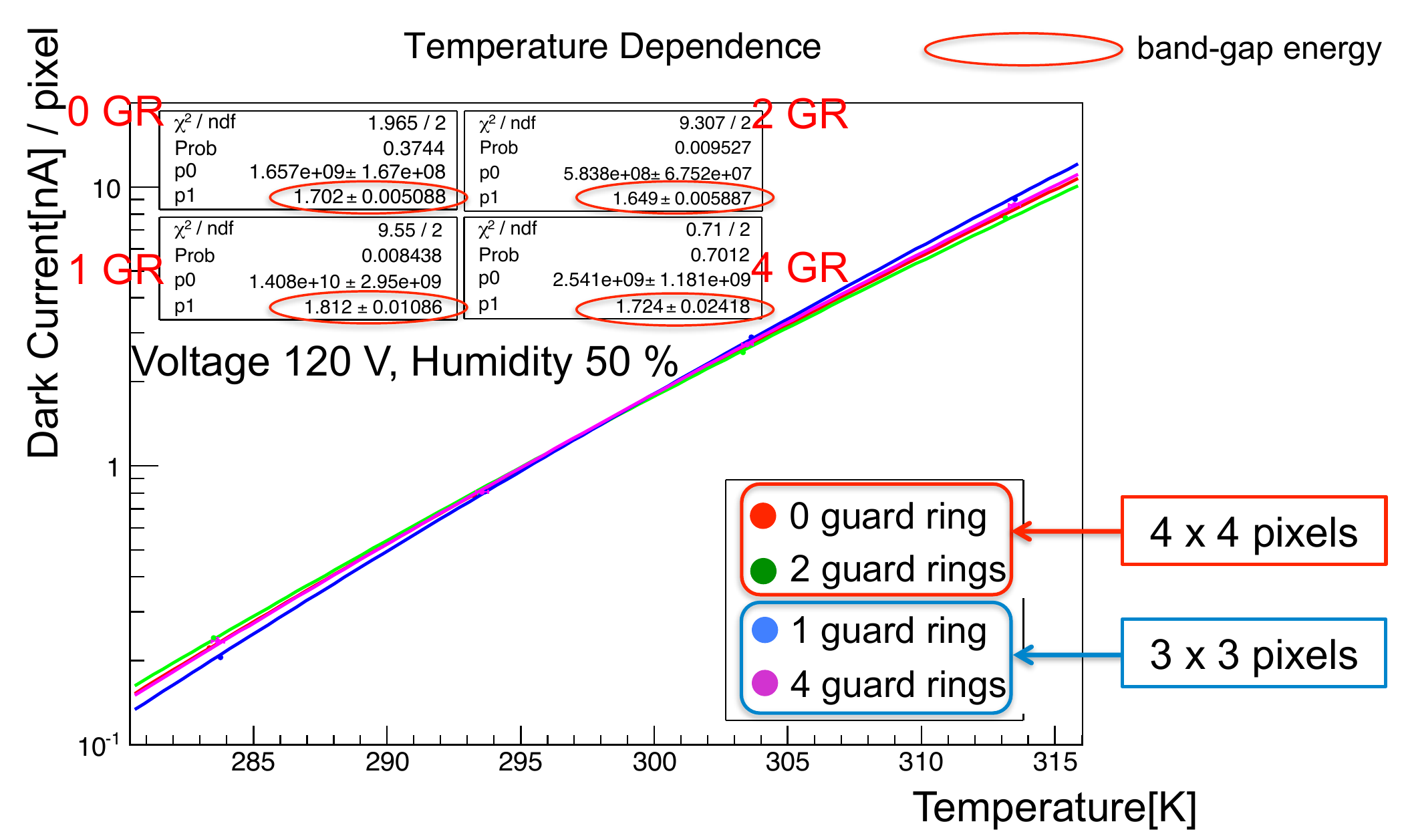}
      \caption{Temperature dependence. Horizontal axis shows temperature, and vertical axis shows dark current normalized by the number of pixels. Fitting parameter of p1 means silicon's band-gap energy.}
      \label{fig11}
    \end{center}
  \end{minipage}
  \begin{minipage}{0.5\hsize}
    \begin{center}
      \includegraphics[width=8.3cm, bb=0 0 475 329]{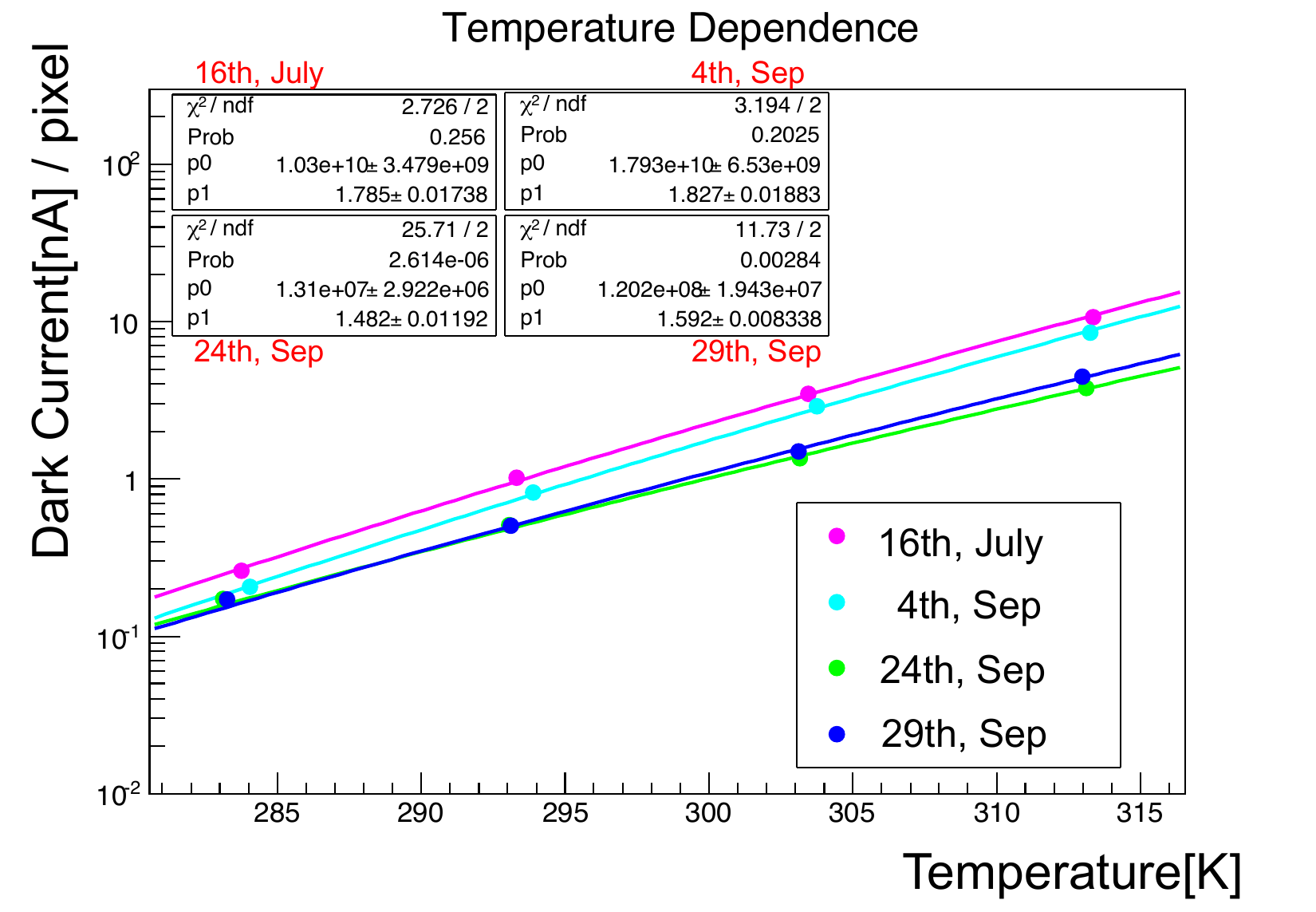}
      \caption{Reproducibility of I-T correlation for 1 guard-ring type. Other types also cannot get reproducibility.}
      \label{fig12}
    \end{center}
  \end{minipage}
\end{figure}
\begin{figure}[htbp]
  \begin{center}
    \includegraphics[width=12cm, bb=0 0 619 369]{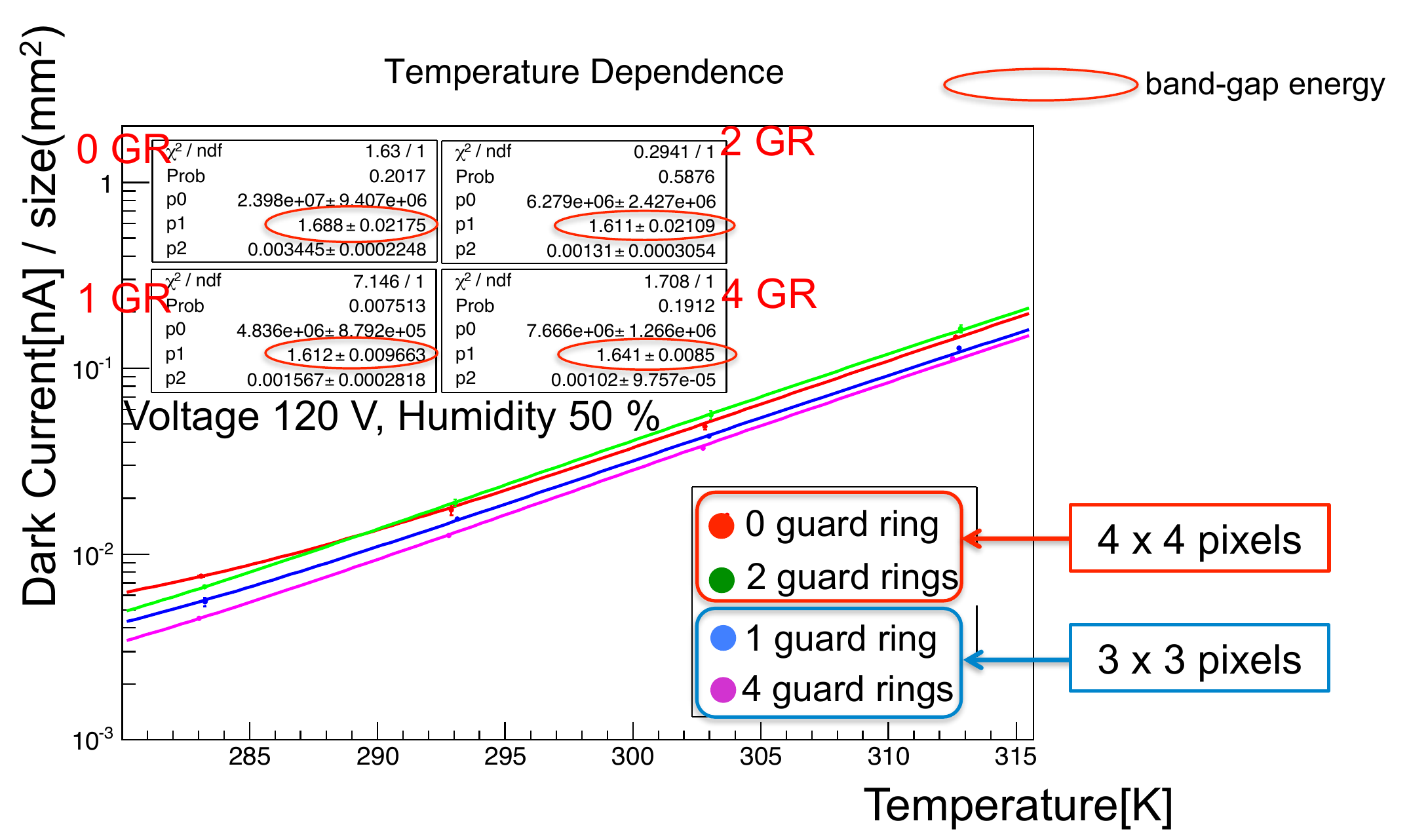}
    \caption{Improved result of temperature dependence on dark current of Si-pad}
    \label{fig13}
  \end{center}
\end{figure}

Figure 14 shows the result of the measurement of laser injection for 0, 1, 2 and 4 guard-ring(s) types.
Percentage shown on the graphs means a slope on the graph for each pixel.
For 1 guard-ring type, ring-formed crosstalk was seen because current flows along the guard-ring.
Uncertainty of the slope is less than 3\% for the same chip.
On the other hand, for the other types of guard-ring samples, there are no significant crosstalks.
For 2 and 4 guard-ring types, guard-rings are divided so that current cannot flow to other channels.
For 0 guard-ring type, current also cannot flow.
\begin{figure}[htbp]
  \begin{center}
    \includegraphics[width=13cm, bb=0 0 562 444]{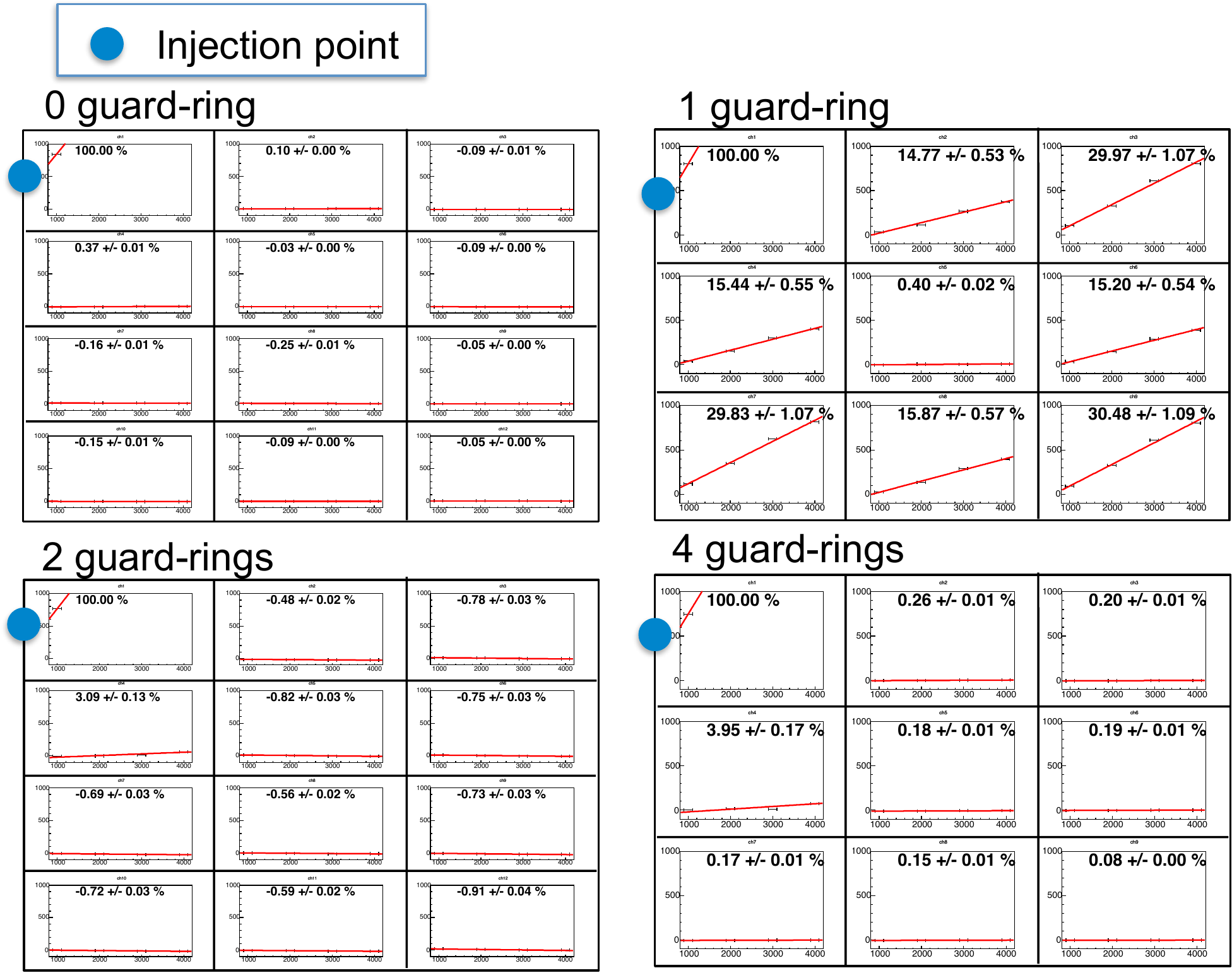}
    \caption{Laser injection for 0, 1, 2 and 4 guard-ring(s) types. Horizontal axis shows response of the nearest pixel to the injection point, so the axis is proportional to laser power. Vertical axis shows the response of each pixel. Value of the slope at each pixel varies chip by chip, but pattern of the slopes was consistent with all chips of the identical specification.}
    \label{fig14}
  \end{center}
\end{figure}
\begin{figure}[htbp] 
  \begin{center}
    \includegraphics[width=7cm, bb=0 0 470 363]{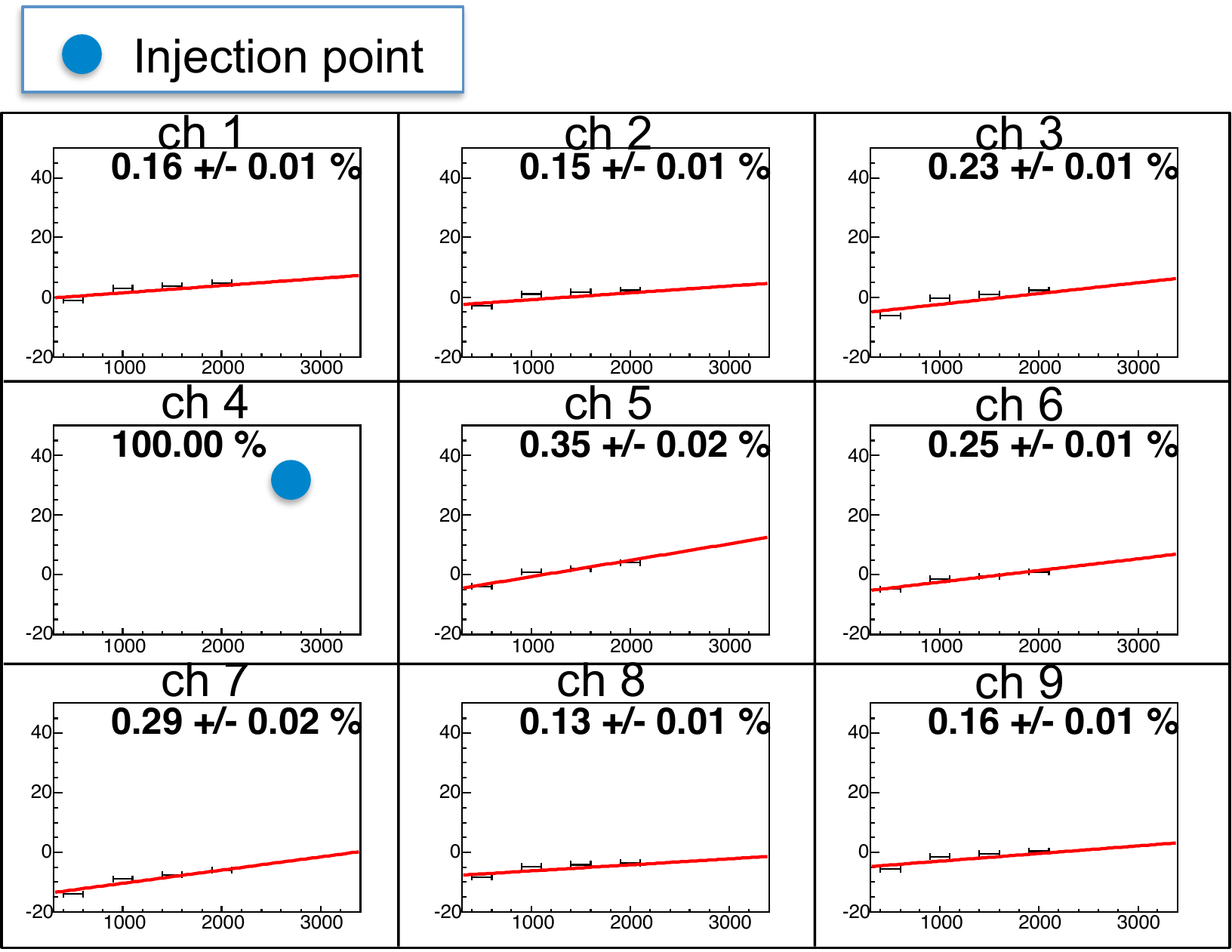}
    \caption{Result of laser injection inside pixel with a meshed chip shown in similar format with Figure 14.}
    \label{fig15}
  \end{center}
\end{figure}                                                                                                                        

Figure 15 shows the result of laser injection to the meshed electrode.
Laser point is shown in Figure 10.
For all channels, crosstalks are less than 0.4 \%.
The difference of slope can be due to the small difference of gaps caused by a small variation at the production. 
We plan to compare between 0 and 1 guard-ring types.

\section{Summary and Plan}
We measured temperature dependence of dark current and responses of laser injection for Si-pad. 
For temperature and humidity dependence of dark current, no significant difference appeared among different guard-ring types.
For laser injection between guard-ring and edge of a pixel, crosstalk is seen at the 1 guard-ring type, but crosstalk is not seen at 0, 2 or 4 guard-ring(s).
For the laser injection inside a pixel, crosstalk between pixels is at around 0.4 percent.
In conclusion, currently we do not see any disadvantages in no guard-ring sensors.
Crosstalk from the guard-ring is only seen in the guard-ring type.

For the next step, we will improve the setup of I-V measurement to reject the edge effect.
We also plan to move to thicker sensors of for example, 500 ${\mu}$m thickness to achieve better energy resolution.
We have to inspect the characteristics of 500 ${\mu}$m sensors to check the no guard-ring sensor does not have disadvantages.
At last, we are preparing to establish measurement procedure for quality control in mass Si-pad production.

\section*{Acknowledgement}
We thank colleagues in LLR and CALICE-Asia groups, especially Dr. Vladislav Balagura for useful discussion.
This work was supported by MEXT / JSPS KAKENHI Grant Numbers 23000002 and 23104007.

\end{document}